\documentclass[preprint,1p,number,sort&compress]{elsarticle}%
\usepackage{color,amsmath,amssymb,amsfonts,natbib}

\newcommand{\beq}{\begin{equation}}
\newcommand{\eeq}{\end{equation}}
\newcommand{\nn}{\nonumber}

\newcommand{\erf}[1]{Eq.~(\ref{#1})}

\newcommand{\srf}[1]{Sec.~\ref{#1}}

\newcommand{\sch}{Schr\"odinger}

\newcommand{\sq}[1]{\left[ {#1} \right]}
\newcommand{\cu}[1]{\left\{ {#1} \right\}}

\newtheorem{theorem}{Theorem}

\newtheorem{assumption}[theorem]{Assumption}

\newtheorem{criterion}[theorem]{Criterion}
\newtheorem{definition}[theorem]{Definition}

\newcommand{\EPR}{\mathcal{EPR}}
\newcommand{\Dis}{\mathcal{D}{\rm ist}}
\newcommand{\Rep}{\mathcal{R}{\rm ep}}
\newcommand{\Pre}{\mathcal{P}{\rm re}}

\newcommand{\Com}{\mathcal{C}{\rm om}}

\newcommand{\Disc}{\mathcal{D}{\rm isc}}

\newcommand{\follows}{\Longleftarrow}
\newcommand{\et}{\wedge}

\definecolor{nred}{rgb}{0.85,0.05,0.0}
\definecolor{nblack}{rgb}{0,0,0}
\definecolor{ngreen}{rgb}{0.2,0.7,0.2}

\newcommand{\blk}{\color{nblack}}
\newcommand{\grn}{}{\color{ngreen}}

\begin{document}

\title{Quantum discord is Bohr's notion of non-mechanical disturbance introduced to 
answer EPR}

\author{Howard M. Wiseman} 
\ead{H.Wiseman@Griffith.edu.au}
\address{
Centre for Quantum Dynamics, Griffith University,  Brisbane, Queensland 4111, Australia \\ \vspace{5ex}
{\rm Accepted for publication in} Annals of Physics {\rm on 4th May 2013}}

\begin{abstract}
By rigorously formalizing the Einstein-Podolsky-Rosen (EPR) argument, and Bohr's reply, one can appreciate 
that both arguments were technically correct. Their opposed conclusions about the completeness of quantum mechanics hinged upon 
an explicit difference in their criteria for when a measurement on Alice's system can be regarded as not disturbing 
Bob's system.  The EPR criteria allow their conclusion --- incompleteness --- to be reached by 
establishing the physical reality of just a single observable $q$ (not of both $q$ and its conjugate observable $p$), 
but I show that Bohr's definition of disturbance prevents the EPR chain of reasoning from establishing even this. 
Moreover,  I show that Bohr's definition is intimately related to the asymmetric concept 
of quantum discord from quantum information theory: if and only if the joint state has no Alice-discord, she 
can measure any observable without disturbing (in Bohr's sense) Bob's system. Discord can be present even when 
systems are unentangled, and this has implications for our understanding of the historical development 
of notions of quantum nonlocality.
\end{abstract} 


\maketitle


\section{Introduction} \label{sec:intro}

No paper is more seminal for the fields of quantum foundations and quantum information 
than the 1935 Einstein, Podolsky and Rosen (EPR) paper \cite{EPR35}. 
It was the first  to discuss the correlations between the measurement results of two 
parties who share an entangled state. Indeed, the term {\em entangled} was 
introduced by \sch\ in the same year \cite{Sch35}, in a paper motivated (as \sch\ explicitly notes) by the 
EPR paper. In another paper of that year \cite{SchPCP35}, \sch\ introduced the term 
`steering' to describe the EPR-phenomenon, a notion which has recently been formalized
in quantum informational terms \cite{WisJonDoh07,JonWisDoh07}, and shown to be a 
sort of quantum nonlocality intermediate between entanglement 
and Bell-nonlocality. This last notion was of course introduced in Bell's 1964 paper \cite{Bel64}. Although it 
sunk Einstein's dream of a local deterministic theory of quantum phenomena, it too was 
directly inspired by the EPR paper, being entitled ``On the Einstein Podolsky Rosen Paradox''.

In this paper, I show that yet another famous paper written in reaction to the EPR paper 
introduced a concept of quantum nonlocality which has only relatively recently been formalized 
(from an independent perspective \cite{HenVed01,OllZur01})  and which is of considerable current interest 
\cite{DakVlaBru10,LanCav10,StrKamBru11}. 
The paper is Bohr's 1935 response \cite{Boh35} to EPR, and the concept is {\em quantum discord}, 
introduced in 2001 \cite{HenVed01,OllZur01}. I am using the term `quantum nonlocality' here 
(and above) in a very broad sense, to refer to any phenomenon requiring  at least two  
parties  who could be arbitrarily separated in space and who must  share a quantum state. 
Indeed, the term has been used (`quantum nonlocality without entanglement' \cite{Ben99-NLWoE}) 
for an effect that is intimately related to quantum discord.

Bohr famously gave his paper the same title as the EPR paper: 
``Can quantum mechanical description of physical reality be considered complete?'' 
But of course while EPR answered this questions `no', Bohr answered `yes'. 
There are different opinions \cite{Mer98,Beller99,Nor05,BruZuk12} 
about whether EPR, or Bohr, made logical errors in this debate, 
or whether (as I advocate here) the disagreement can be understood 
as arising from EPR and Bohr using different definitions. 
Under the latter scenario, their different definitions presumably 
reflect their different conceptions of the nature of the world and our relation to it,  and 
one could still discuss whether Bohr's or Einstein's world view is to be preferred. It is not the purpose of this 
paper to delve into such philosophical matters, but rather to isolate the essence of the disagreement and 
to formalize it. As I will show, the most reasonable formalization of the arguments of EPR and Bohr allow both 
of them to be logically correct, with an {\em explicit} difference in one definition (that of {\em disturbance}) 
being responsible for their different conclusions. 

Quantum discord $\delta$ was defined by Ollivier and Zurek in 2001, as a 
``measure of the quantumness of correlations'' between two parties, Alice and Bob, who share a quantum state \cite{OllZur01}.
 Roughly, the Alice-Discord $\delta(\beta|a)$ quantifies the loss in the amount of quantum information Alice's system  
 has about Bob's system $\beta$ which occurs 
 when Alice makes a projective measurement $a$ on her system. In the same year, Henderson and Vedral considered a similar quantity, 
 without naming it, but required that it be {\em minimized} over all possible (including non-projective) measurements by Alice \cite{HenVed01}.  This is now usually taken as the definition of the discord of a state \cite{DakVlaBru10,LanCav10,StrKamBru11}. 
 I show that the relation of discord to Bohr's 1935 paper is the following. If and only if the discord is zero, 
 Alice can measure any observable on her system in a way that does not disturb (in Bohr's sense) Bob's system.

Because this paper uses a relatively large number of formal concepts, and employs symbolic logical in some parts, it is helpful to introduce abbreviations for many of these concepts. This is done in 
\srf{sec:Defs}, which also contains some preliminary definitions used later in the paper.  In \srf{sec:EPR} I formalize the EPR argument, 
 and also give a simplified version which involves only a single observable $q_\beta$ for Bob's system. Then I formalize Bohr's counter-argument in \srf{sec:Bohr} and show that it works against both the original EPR argument and the simplified version.  In \srf{sec:Ent} I show the relation between the EPR argument and entanglement, while in \srf{sec:Discord} I show how  Bohr's notion of disturbance relates to the notion of quantum discord. Section~\ref{sec:Discussion} concludes with a discussion of the implications of this work, particularly in the light of Bell's theorem. 

\section{Notation and some definitions} \label{sec:Defs}

\subsection{Preliminary notation}

Consider an experiment involving 
two parties, or agents, Alice and Bob. They can freely 
choose between various actions, which can be regarded as controlling their measurement settings. 
Following Bell's notation \cite{Bel64}, I will denote these $a$ and $b$ respectively. 
These settings then yield outcomes $A$ and 
$B$ respectively (Bell's notation). The {\em preparation} $c$ 
(Bell's notation) is the macroscopic event (or series of events) initiating the experiment --- apart from the measurement settings 
controlled by Alice and Bob ---  the details of which  remain fixed between one experimental run and the next. 
For $c$ to have any role in the {\em correlations} between $A$ and $B$, it (or parts of it) must lie in the past light cones of both  
$A$ and $B$. 

To connect to the language of EPR and Bohr, it is necessary to talk of {\em systems}. 
To give this an operational meaning (that is, one grounded in macroscopic events), 
assume that events $a$ and $A$ 
are locatable within a space time region $\alpha$ which is disjoint from another region $\beta$ containing $b$ and $B$. 
We can then identify Alice's and Bob's systems with 
$\alpha$ and $\beta$ respectively. For simplicity I will assume that $c$ lies in the backward light cones of both $\alpha$ and $\beta$, 
which I denote $t(c) < t(\alpha), t(\beta)$. 
For Bell nonlocality, it is often assumed that $\alpha$ is space-like separated from $\beta$, so that it is true neither that 
$t(\alpha) < t(\beta)$ nor that $t(\alpha) > t(\beta)$.  This arrangement was never assumed by EPR or Bohr, but is certainly compatible with the physics they considered. EPR and Bohr  talk about one party (say Alice) making {\em predictions} regarding Bob's system, so it does seem reasonable to require that $t(\alpha) {>} t(\beta)$ {\em not} be true. 

 EPR also refer to {\em physical quantities} pertaining to systems $\alpha$ and $\beta$. I will denote these by $\hat{a} \in \mathbb{P}_\alpha$ and $\hat{b} \in \mathbb{P}_\beta$. I use a hat because in quantum mechanics physical quantities are represented by operators, but there is no implication here that quantum mechanics must be correct. Because there are many ways to measure a physical quantity, $\hat a$ is associated with an equivalence class of settings: $a \in \mathbb{S}_{\hat a}$, and similarly for $\hat b$ and $b$. 
 This equivalence class will be defined formally below. 
 Finally, I also define the set of possible outcomes $A$ for a measurement setting $a \in \mathbb{S}_{\hat a}$ as $\mathbb{O}_{\hat a}$ and similarly for $B$. In the remainder of this paper, unless otherwise specified, the symbol $\forall\ a$ is to be understood as meaning all possible $a$; that is, $\forall\ a \in \mathbb{S}_{\hat a}$ for some $\hat a \in \mathbb{P}_\alpha$;  and likewise for $b$, $A$, and $B$, {\em mutatis mutandis}. 

\subsection{Some definitions}

I can now present some basic definitions using the above events and variables. 
\begin{definition}[Phenomenon] \label{defPhe}
A phenomenon $\phi \equiv (c,\mathbb{P}^\phi_\alpha,\mathbb{P}^\phi_\beta)$ is  defined by the complete set of relative frequencies for the outcomes $A$ and $B$: 
$$
\{ f_\phi(A,B| a, b, c) : A, B, a, b \} \nn
$$
for some preparation $c$ and for all physical quantities $\hat a \in \mathbb{P}^\phi_\alpha \subseteq \mathbb{P}_\alpha$ and $\hat b \in \mathbb{P}^\phi_\beta \subseteq \mathbb{P}_\beta$.
\end{definition}
Here I am using the notation $\{f:g\}$ to mean the set of values of $f$, where $g$ is the set-index, which ranges over all 
allowed values of $g$. I use the term `relative frequency' rather than `probability' to emphasize that a phenomenon is something empirically observed. Of course it may also be theoretically predicted, and in particular I will use the term {\em quantum phenomenon} for a phenomenon that is predicted by quantum mechanics. The set of quantum phenomena will be denoted $Q$. 
An {\em operational theory} is a means of calculating all the relative frequencies $f_\phi(A,B| a, b, c)$ pertaining to given phenomenon $\phi \equiv (c,\mathbb{P}^\phi_\alpha,\mathbb{P}^\phi_\beta)$, and which postulates no variables distinct from the specified macroscopic events. 
The notion of equivalence class of measurement settings is also operational, and may depend upon the preparation $c$:  
\beq \label{eq:defSa}
a \in \mathbb{S}_{\hat a|c} \iff \forall\ a' \in \mathbb{S}_{\hat a|c}, A,  
B, b,  \left[  f_\phi(A,B| a, b, c) = \sum_{A'} f_\phi(A',B| a', b, c)\delta_{A',A} \right]
\eeq
Orthodox quantum theory (OQT) is the example {\em par excellence} of an operational theory, 
\grn at least for a preparation procedure $c$ corresponding to a pure state and 
{\em sharp} measurements $a$ and $b$ (i.e.~with rank-one probability operators \cite{WisMil10}). 
This case is special because it has a unique set of relative frequencies for all observers, as there are no extra 
macroscopic variables that could be known to some observers but not others. \blk 

Next, we need to note that EPR use the terms {\em description of physical reality} and {\em theory} interchangeably. 
That is, they did not use the term {\em theory} to mean operational theory. Rather, they imagined that 
a {\em theory} could involve additional ({\em hidden}) variables, with postulated relations between the variables 
that could not be derived from the \grn preparation $c$ or \blk 
observable relative frequencies alone. For the purpose of the debate between EPR and 
Bohr, we need consider only a single phenomonen $\phi$ (as defined above), and 
hence the class $\Theta_\phi$ of theories describing this phenomenon. Thus, 
\begin{definition}[Theory] \label{DefThy}
A theory $\theta \in \Theta_\phi$ for a phenomenon $\phi$ consists of 
\begin{enumerate}
\item The set $\Lambda_c^\theta$ of values of $\lambda$.
\item A (non-negative) probability meausure $d\mu^\theta_c(\lambda)$ on $\Lambda^\theta_c$.
\item The (non-negative) conditional probabilities $ P_\theta(A,B|a,b, c, \lambda)$
\end{enumerate}
which reproduces the phenomenon for all  $\hat a \in \mathbb{P}^\phi_\alpha, \hat b \in \mathbb{P}^\phi_\beta$:  
$$
\forall\ A, B, a, b, 
\int_{\Lambda^\theta_c} d\mu^\theta_c(\lambda) P_\theta(A,B|a,b, c, \lambda) = f_\phi(A,B|a, b,c)
$$
\end{definition}
EPR's intuition was that the hidden variables $\lambda$ could 
allow  a better {\em description of physical reality} than that given by an operational theory. 

\section{The EPR paper} \label{sec:EPR}

The EPR paper grew out of decades of disquiet Einstein felt about quantum theory, exacerbated by the rise of 
the Copenhagen interpretation as the orthodox one \cite{Nor05}. In this paper I am not concerned with putting the EPR paper in the context of Einstein's earlier or later writings (unlike in \cite{Wis06a}), but rather with presenting the formal argument in the EPR paper itself 
quite rigorously  (unlike in \cite{Wis06a}).  
The reason, as I explained in Sec.~\ref{sec:intro}, is that it was at the level of formal argument that Bohr responded to the EPR paper, and it is at this level that the connection to discord lies.

\subsection{The EPR criteria} \label{sec:EPRdefs}

The EPR paper is of course concerned with completeness and, as with almost all concepts they introduce, they explain it reasonably precisely:
\begin{quote}
Whatever the meaning assigned to the term {\em complete}, the following requirement 
for a complete theory {seems} to be a necessary one: {\em every element of the physical reality must have a counterpart in the theory}.
\end{quote}
Here, and --- except where noted --- below, the italics are as in the original. This criterion can be expressed using formal logic as 
\begin{criterion}[Completeness] \label{ComCriterion}
$$\forall \theta \in \Theta_\phi, \ \Com(\theta)  \implies \sq{ \forall\  \hat b \in \mathbb{P}^\phi_\beta , \ \EPR(\hat b|c) \implies  \Rep_\theta(\hat b|c) }.$$
\end{criterion} 
Here I have used $\Rep_\theta(\hat b|c)$, standing for `is represented', to denote EPR's concept that $\hat b$ `has a counterpart' in theory $\theta$, and $\EPR(\hat b|c)$ to mean that the property $\hat b$ is an element of physical reality. 
 The notation used in the above requires some comment. First,  
I use the notation  $\EPR(\hat b|c)$, rather than $\EPR_\theta(\hat b|c)$ because EPR are quite clear that 
 their sufficient condition (see below) for something to be an element of physical reality 
must come from the phenomenon, not the theory:  
\begin{quote}
The elements of physical reality cannot be determined by {\em a priori} philosophical considerations, but must be found by an appeal to  results of experiments and measurements. 
\end{quote}
Second, the notation $\EPR(\hat b|c)$, rather than $\EPR(\hat b|a,c)$ should not be read to imply that the reality of a physical property of system $\beta$ is necessarily independent of the process of measurements performed upon system $\alpha$. However, as I will discuss below, the sufficient criterion EPR use for $\EPR(\hat b|c)$ applies only in the case where the measurement $a$ does not disturb the system $\beta$. In this case, they are again clear that 
\begin{quote} 
No reasonable definition of reality could be expected to permit [that] the reality of [properties of the second system] depend upon the process of measurements carried out on the first system, which does not disturb the second system in any way.
\end{quote}
Thus in the case of interest, it must be the case that $\EPR(\hat b|c) = \EPR(\hat b|a,c)$. Third, EPR's sufficient criterion (below) implies that it would be redundant to consider $\EPR(\hat b|b,c)$. It would not, however, be sensible to consider 
$\EPR(\hat b|B,b,c)$, as this would make EPR's requirement of being able to predict the outcome ($B$) of the measurement  
completely trivial. Since we can thus restrict to considering $\EPR(\hat b|c)$, it follows that we must restrict the conditionals 
to what `has a counterpart' in theory $\theta$ in the same way. \grn Thus we obtain a local concept of representation, $\Rep_\theta(\hat b|c)$, and thus a concept of local elements of physical reality, which is critical to EPR's argument \cite{Nor11}. \blk

EPR do not actually define what it means for $\hat b$ to `have a counterpart in the theory', but I think it is uncontroversial to take it to imply that the outcome of any measurement of $\hat b$ is determined in the theory:
\begin{criterion}[Representation in the theory] \label{RepDef}
$$\Rep_\theta(\hat b|c) \implies \forall\ B, b, \lambda, \ P_\theta(B|b,\lambda,c) \in \cu{0,1}. $$
\end{criterion}
Regarding elements of physical reality, EPR stress that ``[a] comprehensive definition of reality is \ldots unnecessary for our purpose.'' Instead, they give the following sufficient criterion: 
\begin{quote} 
 {\em If, without in any way disturbing a system, we can predict with certainty (i.e., with probability equal to unity) the value of a physical quantity, then there exists an element  of physical reality corresponding to this physical quantity.} 
\end{quote}
To enable a non-disturbing measurement on $\beta$ they consider an {\em indirect} measurement (i.e. a measurement 
on another system $\alpha$), so formally  we have:  
\begin{criterion}[Element of Physical Reality]\label{EPRcriterion}
$$\EPR(\hat b|c) \follows \exists\  a :  [\ \neg \Dis(\beta|a,c) \et \Pre(\hat b|a,c) \ ].$$ 
\end{criterion}
Here $\Dis(\beta|a,c)$ means that the measurement $a$ disturbs the system $\beta$, while $\Pre(\hat b|a,c)$ 
means that the measurement $a$ makes the value of $\hat b$ predictable.  
Like representation, predictability\footnote{Note that it is important to distinguish between  predictability and determinism; 
see e.g.~Ref.~\cite{CavWis12}.} can  be defined uncontroversially: 
\begin{definition}[Predictability]
$$\Pre(\hat b|a,c) \iff \forall\  B, b,  A, \ f(B|A,a,b,c) \in \cu{0,1}.$$
\end{definition} 
Note the use here of the operationally determined relative frequencies $f$, not the theory probabilities $P_\theta$.
The second new notion --- disturbance --- requires clarification. EPR say (with my emphasis):
\begin{quote}
A definite value of the coordinate [$x$], for a particle in the state given by Eq.~(2) [$\psi(x)=e^{(2\pi i / h)p_0x}$], is thus not predictable, but may be obtained only by direct measurement. Such a measurement however {\em disturbs} the particle and thus alters its state. 
 After the coordinate is determined, the particle will no longer be in the state given by Eq.~(2). 
\end{quote} 
Here by `state' EPR obviously mean the particle's quantum state --- its state according to OQT 
\footnote{ On the basis of other passages in the EPR paper, a case can be made that EPR's notion of disturbance 
was meant to apply to the `real situation' (to use one of Einstein's later phrases) of a system, not just its quantum state 
[R.~Spekkens, private communication (2012)]. One could formalize this by introducing a theory-dependence 
in the criterion for non-disturbance [i.e.~$\neg\Dis_\theta(\beta|a,c)$] by using the theoretical probabilities
$P_\theta$ rather than the relative frequencies $f$. However one could not simply substitute this for $\Dis(\beta|a,c)$ in 
Criterion \ref{EPRcriterion} because that criterion uses, besides the disturbance which is in question, 
only operational concepts, namely $\EPR(\hat b|c)$ and $\Pre(\hat b|a,c)$. However one could 
replace $\neg\Dis(\beta|a,c)$ in Criterion \ref{EPRcriterion} by $\exists \ \theta :  \neg\Dis_\theta(\beta|a,c)$, 
since every theory must reproduce the operational predictions, so if there is no disturbance in some theory 
then there will be no disturbance in the operational sense (the criterion I have advocated).  \grn 
Making this change would make no essential change to my interpretation of EPR and Bohr because 
EPR do not justify the lack of disturbance from examining the lack of change in either the 
relative frequencies, or the theoretical probabilities. Rather, they justify it from the fact that ``the two systems 
no longer interact,'' as discussed in the main text.\blk}.
We can avoid referring to quantum theory (as has been the case with my definitions so far) by recognizing that
the state of system $\beta$ in OQT can be identified with a list of the possible relative frequencies of results $B$ 
that will be obtained upon performing measurements $b$. For a measurement $a$ to 
 `alter the state' means that at least one relative frequency is affected by choosing $a$ rather than some other
 measurement $a'$ (which could be the null measurement).  Since EPR imply that the alteration of the state follows 
 from the disturbance of the system (the `particle'), it seems reasonable to formalize their notion of non-disturbance 
 as the sufficient criterion  
 \begin{criterion}[No Disturbance \`a la EPR] \label{defDis}
$$\neg \Dis^{\rm EPR}(\beta|a,c)  \follows  \forall\  a',B, b, \ f(B|a,b,c) =  f(B|a',b,c).$$
\end{criterion} 
Note the use of the EPR superscript here because later I will consider a different definition for disturbance, 
that proposed by Bohr. If $\alpha$ and $\beta$ are space-like separate, then no-disturbance  
follows from Criterion~\ref{defDis} and the no-superluminal-signalling (from Alice to Bob) assumption. \grn 
EPR themselves make 
the stronger assumption, that ``since at the time of measurement the two systems no longer interact, 
no real change can take place in the second system in consequence of anything that may be done to the first system.'' 
Presuming that `no real change' must mean no disturbance (see again the preceding footnote), it follows that EPR embraced the following 
\begin{assumption} Presuming `$\alpha$ and $\beta$ no longer interact' $\implies \forall a, \ \neg \Dis^{\rm EPR}(\beta|a,c)$. \label{ass}
\end{assumption}
Thus if one disagreed with the operational Criterion \ref{defDis} I have abstracted 
from the EPR paper, one could replace it, without significantly affecting any of my analysis or conclusions, 
 with Assumption \ref{ass}. 

\blk 

\subsection{The EPR theorem} \label{sec:EPRthm}

Having suitably formalized EPR's definitions and criteria, we can now turn to what they prove. 
It will be apparent that the two EPR criteria, for completeness and for elements of physical reality, 
being necessary and sufficient respectively, are the minimum required for proving the EPR theorem.
By contrast, the logical argument EPR employ, while correct, is unnecessary complicated. 
This must have contributed 
 to Einstein's complaint that the paper, apparently written for publication by Podolsky, was 
 `smothered in formalism' \cite{Fin96}\footnote{The formal definitions and criteria I have used may also have
 been unpalatable to Einstein, but I think are necessary to best appreciate Bohr's argument. \grn After all, Bohr wrote in 
 response to the EPR paper, not to any other, less formal but arguably clearer, statements by Einstein.\blk}.
Here I present a simpler, but logically equivalent, argument. 

Consider the quantum phenomenon $\phi$ in which $c$ is the preparation of a two-particle entangled 
state,  such that at some time $t_1$, $\hat q_\alpha = \hat q_\beta - x_0$, and $\hat p_\alpha = - \hat p_\beta$.  
Here the subscript $\alpha$ ($\beta$) denotes the particle present in Alice's (Bob's) lab at the time of measurement, 
which I take to be $t_1$ in both cases. Following EPR, take the particles to be {\em `no longer interacting'}, 
which will be the case if Alice's and Bob's labs are spatially separate. 
Under these conditions it follows that 
(i) $\Pre(\hat q_\beta|q_\alpha,c)$, because if Alice measures $q_\alpha$ with result $Q_\alpha$, 
she can predict that   if Bob 
were to measure  $q_\beta$ he would find result $Q_\beta = Q_\alpha + x_0$. It is also the case that (ii) 
$\neg \Dis^{\rm EPR}(\beta|\hat q_\alpha,c)$, because there is no signalling from Alice to Bob. 
From Criterion~\ref{EPRcriterion} it follows that (iii) $\EPR(\hat q_\beta|c)$. 
By an identical argument, it follows that (iv) $\EPR(\hat p_\beta|c)$. 

Now assume that (v) OQT is complete. Here I regard OQT, for a particular phenomenon $\phi$, 
as actually a set $O_\phi \subset \Theta_\phi$ of theories  each of 
which uses only the concepts 
of OQT (quantum states and operations). There might be some argument as to whether 
EPR's notion of completeness should be expected to hold when dealing with mixed states, but in the present 
situation the EPR correlations defined above can pertain only for a pure state, and indeed there 
is only one theory $\theta \in O_\phi$ which yields these correlations. Thus a defender of the completeness
of $O_\phi$ must take this $\theta$ to be complete. Then, from Criterion~\ref{ComCriterion}, 
and results (iii) and (iv) above, it follows that (vi) for this $\theta$, 
$\Rep_{\theta}(\hat q_\beta|c)$ and $\Rep_{\theta}(\hat p_\beta|c)$.  By Criterion~\ref{RepDef}, 
this means that (vii) the theory $\theta$ determines both the value of $\hat q_\beta$ and the value of 
$\hat p_\beta$. However, (vii) is obviously false because in OQT a pair of non-commuting 
operators do not have determined values under any circumstances. Thus the assumption of completeness (v) is false: 
\begin{theorem}[EPR, 1935] $$\exists\   \phi \in Q : \forall\ \theta \in O_\phi, \ 
\neg \Com(\theta).$$ \label{thm:EPR}
\end{theorem}
Here $Q$ is the set of all quantum phenomena. 

As has been noted many times before --- e.g. Refs.~\cite{Red87,Fin96,Nor05} ---  the EPR argument can be made even simpler. 
Given the EPR correlations, 
and hence the unique OQT $\theta$ which reproduces them, we do not need to consider 
a pair of observables at all. It is enough to assume completeness, and derive $\Rep_{\theta}(\hat q_\beta|c)$, 
for example. To obtain a contradiction, simply note that, in the EPR state, $\hat q_\beta$ does not 
have a determined value (it in fact has an infinite uncertainty). If one does not assume that QM is correct, 
then there is a motivation not to 
discard the non-commuting variables version \cite{Wis06a}, but for the purposes of this paper 
 the one-variable argument is just as good. Indeed, as we will see in Sec.~\ref{sec:reactBohr}, 
the simpler argument helps to highlight the generality of Bohr's counter-argument. 

\section{Bohr's Response} \label{sec:Bohr}

\subsection{Formalizing Bohr's counter-argument} 

Bohr's response \cite{Boh35} to EPR's incompleteness theorem is notoriously hard to follow \cite{Beller99,Wis06a,How07}.   
In the greater part of his reply, Bohr did not address the EPR entangled state, but rather discussed the 
complementarity of properties of a single particle, as he had previously done to defend the 
 {\em consistency} of OQT against Einstein's earlier attacks. 
This defence was based upon the mechanical disturbance of the quantum 
system by the apparatus, which was not obviously relevant to the EPR scenario since the incompleteness 
lies in the quantum description of Bob's system, which need not interact with any apparatus.  
Bohr himself, when reviewing the Einstein--Bohr debates more than a decade later \cite{Bohr49}, 
contrasted the `lucidity' of the EPR paper \footnote{It must however be borne in mind that Bohr was writing this 
summary for a volume in honour of Einstein's 70th birthday.} with his own ``inefficiency of expression which must have made it very difficult to appreciate the trend of the argumentation.'' 

Despite this, I now believe (contrary to \cite{Wis06a}) that Bohr's counter-argument can be interpreted as a precise rejection of one of EPR's criteria, and that Bohr guides the reader to this in his later review. There, Bohr again refers back to his earlier defences of complementarity, summing up  \cite{Bohr49}: ``As repeatedly stressed, the principal point is here that such measurements [of complementary properties] demand mutually exclusive experimental arrangements.'' He then goes on to quote from his 1935 paper (truncated here by me): 
\begin{quote}
From our point of view we now see that the wording of the above-mentioned criterion of physical reality proposed by Einstein, Podolsky and Rosen contains an ambiguity as regards the meaning of the expression `without in any way disturbing a system.' Of course there is in a case like that just considered no question of a mechanical disturbance of the system under investigation during the last critical stage of the measuring procedure. But even at this stage there is essentially the question of {\em an influence on the very conditions which define the possible types of predictions regarding the future behaviour of the system.} Since these conditions constitute an inherent element of the description of any phenomenon to which the term Ôphysical realityÕ can be properly attached, we see that the argumentation of the mentioned authors does not justify their conclusion that quantum-mechanical description is essentially incomplete.
\end{quote}
With the word `ambiguity' here, Bohr heralds that there are two possible ways way to understand `disturbing a system'. 
In the short note in Nature \cite{Boh35a} announcing his forthcoming reply to EPR, in the first sentence of the abstract of that reply \cite{Boh35}, 
and in its introduction, Bohr identifies the supposed weakness of EPR's argument 
as being an `essential ambiguity' in their criterion for physical reality. This again shows that the `ambiguity' passage is key to understanding 
his rebuttal of EPR. 

Bohr seems, quite reasonably, to associate the EPR notion of disturbance, Criterion \ref{defDis} (or Assumption \ref{ass}), 
with `mechanical disturbance' and concedes that there is certainly no disturbance of that sort on Bob's system in the EPR scenario. But implicitly he rejects EPR's idea of disturbance as too narrow, and proposes a broader notion: 
a system $\beta$ is disturbed by a measurement $a$ if that measurement  
influences the ``conditions which define the possible types of predictions regarding the future behaviour of the system.''
What is it that defines these possible predictions by Alice regarding Bob's system $\beta$? 
Nothing but the joint quantum state, modulo any reversible transformations on system $\alpha$. 
In operational (non specifically quantum) terms, $a_1$ is non-disturbing if and 
only if choosing to perform $a_1$ --- rather than $a'$, say --- does not preclude the possibility of obtaining,  via some {\em subsequent} measurement $a_2$, the same information about $\beta$ as would have been yielded by the measurement $a'$:
Of course we also want to avoid mechanical disturbance, so we also require that $a_1$ be non-disturbing in the EPR sense, 
and likewise for $a_2$ and $a'$. Therefore for simplicity in the below I assume that there is no signalling possible from $\alpha$ to $\beta$ so that all $a$-type measurements are non-disturbing on $\beta$ in the EPR sense. 
This allows Bohr's notion of non-disturbance to be formalized as  follows (where $\delta$ denotes the Kronecker $\delta$-function)
\begin{definition}[No Disturbance \`a la Bohr] \label{defDis2}
\begin{align} \neg \Dis^{\rm Bohr}(\beta|a_1,c) \iff  & \forall a', A_1, \ \exists \ a_2  :  [t(a_2) > t(A_1)] \ \et  \nn \\ 
& \sq{\forall A' , B, b, 
\   f(B|A',a',b,c) = \sum_{A_2,A_1}f(B|A_2,A_1,a_2,a_1,b,c) \delta_{A',A_2}}  . \nn
\end{align}
\end{definition}

Some comments are in order. First, just as with the EPR notion of disturbance, I have formulated Bohr's notion  
in an operational way, with no reference to quantum mechanics or any other theory. Second, in saying that a suitable $a_2$ exists for all $A_1$ I am allowing for an {\em adaptive} measurement \cite{WisMil10}, in which the choice of the second measurement $a_2$ may depend upon the outcome $A_1$ of the first measurement $a_1$. Third, I have been a bit loose with notation in using $f(B|A_2,A_1,a_2,a_1,b,c)$, when my Definition \ref{defPhe} allowed only one measurement for each party. I trust the reader can understand the generalization implied by the new notation here. Fourth, I again stress that to avoid overly long expressions, the above is to be understood to apply only when  $\neg\Dis^{\rm EPR}(\beta|a,c)$ for $a \in \{a_1,a_2,a'\}$.

Bohr does not formally explain how his concept of disturbance impacts upon EPR's notion of physical reality, but it is easy to do so. The EPR argument relies upon $\neg\Dis(\beta|q_\alpha,c)$ being true. This is the case by EPR's concept of disturbance, but not by Bohr's, as can be seen as follows. Consider, in the EPR scenario, if Alice were to measure $p_\alpha$ instead of $q_\alpha$. This would give the conditional probability $f(P_\beta|P_\alpha,p_\alpha,p_\beta,c) = \delta(P_\beta+P_\alpha)$. But if Alice does measure $q_\alpha$ then 
this projects Bob's state into a $\hat q_\beta$-eigenstate, with infinite momentum uncertainty. There is nothing Alice can do in her lab anymore to enable her to gain any information about $p_\beta$; there is no subsequent measurement $a_2$ she can perform 
such $f(P_\beta|A_2,Q_\alpha,a_2,q_\alpha,p_\beta,c) = \delta(P_\beta+ A_2)$.
Thus from Definition \ref{defDis2}, $\neg\Dis^{\rm Bohr}(\beta|q_\alpha,c)$ is false. 
Consequently, with Bohr's notion of disturbance, $\hat q_\beta$ is not an element of physical reality, and the EPR incompleteness proof fails. 

\subsection{Further commentary} \label{sec:reactBohr}

Bohr's response explicitly rejects  EPR's Assumption \ref{ass} 
that with non-interacting systems a measurement on one should not disturb the other. One could even formalize Bohr's 
conviction that the debate centred on an `essential ambuguity'   as a theorem:
\begin{theorem}[Bohr, 1935] \label{thm:Bohr}
$$\exists\   \phi \in Q : \exists \ a :  \ \neg\Dis^{\rm EPR}(\beta|a,c) \et \Dis^{\rm Bohr}(\beta|a,c).$$
\end{theorem}
One could also epitomize Bohr's response in terms drawing upon (but inimical to) EPR: 
\begin{quote}
Any rational\footnote{Here I have replaced EPR's `reasonable' (which is perhaps best understood to mean `locally causal' \cite{Nor11})
by `rational' since Bohr states \cite{Boh35} that EPR's argument ``discloses \ldots 
an inadequacy of the customary viewpoint of natural philosophy for a rational account of [quantum] phenomena.'' } definition of reality must be expected to permit that the reality of properties of the second system may depend upon the process of measurements carried out on the first system, even when it does not disturb the second system in any {\em mechanical} way. 
\end{quote}

\grn It has been argued in Ref.~\cite{BarRudSpe12} (p.~9), that Bohr was a radical positivist,  
identifying reality with what can be known. My analysis of his position certainly seems compatible 
with that conclusion. The authors of Ref.~\cite{BarRudSpe12} 
further maintain that Bohr's argument against EPR boils down to \blk precisely 
that which EPR had anticipated \cite{EPR35}, 
\begin{quote}
One could object to this conclusion on the grounds that our criterion of reality is not sufficiently restrictive. Indeed, one would not arrive at our conclusion if one insisted that two or more quantities can be regarded as simultaneous elements of reality only when they can be simultaneously measured or predicted.
\end{quote} \grn 
My analysis, by contrast, shows that the point on which he disputes with EPR is actually more subtle, 
which can be best appreciated as follows. \blk As noted in the last paragraph of Sec.~\ref{sec:EPR}, the EPR definitions and criteria do not 
actually require one to consider the simultaneous existence of $q_\beta$ and $p_\beta$ as elements
 of physical reality in order to obtain a contradiction with the completeness of OQT from the 
 EPR correlations. Deriving 
 $\EPR(q^\beta)$ alone is sufficient to obtain the contradiction. Thus if the thrust of 
 Bohr's counter-argument had been to establish that  $\EPR(q_\beta) \et \EPR(p_\beta)$ cannot be true, 
 it would not work against the simplified EPR argument presented at the end of Sec.~\ref{sec:EPR}.
 Luckily for Bohr, his actual argument, based on replacing EPR's Criterion \ref{defDis} 
 of disturbance with his own Definition \ref{defDis2}, is sufficient to prevent the derivation 
 of $\EPR(q_\beta)$ in itself, as noted above. 

In Einstein's considered response to Bohr more than a decade later \cite{Ein49}, he 
implicitly rejected Bohr's notion of disturbance in saying 
\begin{quote}
One can escape from this conclusion [that orthodox quantum theory is incomplete] only by either assuming that the measurement of [$\alpha$] (telepathically) changes the real situation of [$\beta$] or by denying independent real situations as such to things which are spatially separated from each other. Both alternatives appear to me equally unacceptable.
\end{quote}
Bohr's equating of a disturbance of $\beta$ with a change in what Alice could {\em find out} about $\beta$ via measurement of $\alpha$, 
would seem to be an instance of the second of Einstein's unacceptable alternatives, namely `denying [an] independent real situation' to $\beta$. It must be noted however that this denial by Bohr was not special to a scenario with two spatially separated systems; 
Bohr seems to have denied altogether that quantum things had `real situations' in Einstein's sense.  \sch, by contrast, who clave to an  
intrepretation of the wavefunction as both realistic and complete, was compelled to accept the first of Einstein's loopholes, 
the `telepathic change' which \sch\ called `steering' \cite{SchPCP35}, as a ``necessary and indispensable feature'' of QM. 
 But he found this `repugnant' and hoped that QM would be found incorrect in its prediction of steering \cite{SchPCP36}, a hope which only recently has been conclusively dashed by experiment \cite{Wittman12}. 
\blk 

\section{EPR and Entanglement} \label{sec:Ent}

EPR's proof (using their notion of disturbance) of the incompleteness of OQT relies 
critically on the existence of entanglement. Indeed,  for any pure entangled state a proof similar to EPR's can easily 
be constructed if one forgoes the pair of complementary observables (as at the end of Sec.~\ref{sec:EPRthm}), 
and considers the observables $\hat a$ and $\hat b$ whose eigenstates define the Schmidt basis \cite{Nielsen00a} for the entangled state. These are the observables such that a measurement of $\hat a$ enables the results of a measurement 
of $\hat b$ to be predicted (and {\em vice versa}). In all of the early papers in quantum foundations, the states 
considered are pure. Nevertheless it is also easy to prove that even allowing for mixture, an unentangled state cannot possibly 
be used to prove incompleteness from EPR's axioms, as I now show.

 An unentangled quantum state  (an operator on the tensor product of 
the $\alpha$ Hilbert space by the $\beta$ Hilbert space) by definition can be written as 
\beq \label{unent}
\varrho_c = \sum_k \wp_c(k) \ \rho^\alpha_k \otimes \rho^\beta_k,
\eeq
where $\wp_c(k)$ is a probability distribution and $\rho^\alpha_k$ and $\rho^\beta_k$ are also quantum states.  
In this case, a measurement on system $\alpha$ will update the state of system $\beta$ in a purely classical way, from 
$\sum_k \wp_c(k) \rho^\beta_k$ to $\sum_k \wp(k|A,a)\rho^\beta_k$, where $\wp(k|A,a)$ is determined 
via Bayes theorem from $\wp_c(k)$ and the properties of $\hat a$ and $\rho^\alpha_k$.  

Now if the EPR argument is to work there must be observables such that a measurement of $\hat a$ enables the results of a measurement of $\hat b$ to be predicted.  This will be possible only if there is a decomposition of the form (\ref{unent}) 
where each $\rho^\beta_k$ is an eigenoperator of $\hat b$; that is, $\hat b \rho^\beta_k = B(k) \rho^\beta_k$.  
Then, if and only if $\rho^\alpha_k \rho^\alpha_{k'} = 0$ whenever $B(k) \neq B(k')$, there exists a measurement  
$a$ that allows Alice to predict $\hat b$ perfectly. In this case, one  
can derive, following the EPR argument, the result $\EPR(\hat b|c)$. 
However, unlike in the pure entangled case, there is,  even for $\hat b$ nontrivial,  
no contradiction with the assumption of completeness: there is a $\theta \in O_\phi$ in which 
$\hat b$ is represented. It is the version of OQT which interprets the index $k$ in \erf{unent} as 
a variable in the theory, a variable like $\lambda$ in the Definition \ref{DefThy}, 
with $\int_{\Lambda_c^\theta} d\mu_c^\theta(\lambda) \to \sum_k \wp_c(k)$.
This version of OQT is the one which takes the appropriate mixture (\ref{unent}) to be a {\em proper} mixture \cite{dEs89}. 
Then, from the Definition \ref{RepDef}, 
 $\Rep_\theta(\hat b|c)$ is true, since $\hat b$ has a definite value given $k$, namely $B(k)$.  
 
 Of course there are other OQTs,  in particular the one in which there are no variables in the theory apart from the 
 $\varrho_c$ which is uniquely specified by the preparation $c$, so that  
 all decompositions (\ref{unent}) are regarded as {\em improper} mixtures \cite{dEs89}. 
 This particular $\theta \in O^\phi$ 
 {\em is} provably incomplete 
 according to EPR's definition of disturbance. Moreover, adopting Bohr's 
 definition of disturbance does {\em not}  necessarily save the day in this case. 
 That is, there are cases where Bohr would also be forced to say that there is no disturbance,  
 as I will discuss at the end of the following section.  
Thus, unless
 one believes that Bohr would have allowed that OQT is incomplete, one must accept that 
 the question of incompleteness of OQT should be answered negatively 
 only if {\em every}  $\theta \in O_\phi$ is incomplete \footnote{The situation is similar to that 
 in classical physics. There are some classical physical theories, such as 
 thermodynamics, that are incomplete in that they do not postulate any microscopic variables. But that does 
 not mean that classical physics {\em per se} is incomplete.}. This incompleteness for all versions OQT is 
 of course what EPR did prove 
 (again according to their definition of disturbance), but what  we have seen here is that 
 entanglement is necessary (and, if pure, sufficient) for the EPR Theorem \ref{thm:EPR}.

\section{Bohr and Discord}  \label{sec:Discord}

Given the discussion in the preceding section on the criticality of entanglement for the EPR theorem, 
it might be expected that entanglement is equally critical for Theorem \ref{thm:Bohr} (which I have 
called Bohr's theorem), that 
there are quantum phenomena in which even though mechanical disturbance is excluded, 
a measurement on $\alpha$ can still Bohr-disturb system $\beta$. Surprisingly, this is not the case. 
Rather, it turns out that what is critical to Bohr's theorem is the existence of {\em quantum discord}.  
Note that in this section I am concerned with Bohr's non-mechanical disturbance in itself, not its 
role in his counter-agument to EPR,  which has already been treated in Sec.~\ref{sec:Bohr}. I will return 
to the EPR-Bohr debate in Sec.~\ref{sec:Discussion}\blk 

As explained briefly in the introduction, discord was a term introduced by Ollivier and Zurek \cite{OllZur01} in 2001 
as a quantitative property of a quantum state $\varrho_c$ shared between two parties (say Alice and Bob), only 
one of which (say Alice) may perform a measurement $a$. They defined Alice-discord as
\beq \label{defdisa}
\delta(\beta | a , c) = I(\beta:\alpha|c) - J(\beta|a,c),
\eeq
where  $I$ is the quantum mutual information between Alice's quantum system $\alpha$ and Bob's $\beta$, 
and $J$ is the average information Alice has about $\beta$ from the result of her measurement $a$ 
(and her knowledge that the original shared state was $\varrho_c$). 
In classical information theory Alice's measurement could simply reveal the configuration of her system in which 
case the difference $\delta$ between would be zero by definition. In quantum information theory this is not the case, and for some states 
$\varrho_c$, the Alice-discord $\delta(\beta | a, c)$ is nonzero for all possible measurements. Henderson and Vedral 
had previously introduced this quantity without naming it, and considered its minimum over all possible 
measurements, which I will denote $\delta(\beta|c)$. 

The proofs below do not use the magnitude of discord, only whether it is nonzero, which I notate by 
\begin{definition}[Discord]
$$\Disc(\beta |  c) \iff \delta(\beta |  c) > 0.$$
\end{definition}
Now it is well known that $\Disc(\beta | c)$ is false if and only if
\beq \label{nodisc}
\varrho_c = \sum_j \wp_c(j) \ \pi^\alpha_j \otimes \rho^\beta_j ,
\eeq
where $\wp_c(j)$ is a probability distribution, the $\pi^\alpha_j$s are mutually orthogonal rank-one projectors, 
and the $\rho^\beta_j$s are simply normalized states with no other restrictions. 
This is of course an unentangled state, so from \srf{sec:Ent} any measurement by Alice on such a state would 
result in classical updating of her knowledge of Bob's state. We now show how Alice-discord is intimately related to Bohr's notion of disturbance, in that 
\beq \label{nfr}
\forall \ \phi \in Q, \ \left[\neg\Disc(\beta| c) \iff \forall \ \hat{a}, \exists\ a \in {\mathbb S}_{\hat a|c} : \ \neg\Dis^{\rm Bohr}(\beta|a,c) \right]. \eeq
 
 To prove the forward implication, assume that $\Disc(\beta | c)$ is false. 
 Then one way in which Alice can measure an observable 
 $\hat a_1$ is as follows. 
First she measures the observable $\hat a_c$ for which $\pi^\alpha_j$ are eigenoperators. 
This would update her knowledge of Bob's state to $\rho^\beta_j$ for some $j$,  and, if done by projection, 
would leave her system in the corresponding state $\pi^\alpha_j$. On average, 
(that is, ignoring the result $A_c$ which stores the value $j$) this measurement has no affect on the 
state. Thus Alice can then measure $\hat a_1$ by any other means as if she had never measured $\hat a_c$, 
and update her knowledge of $\beta$ thereby. But then recalling the result $A_c$ she can restore the 
initial state of the system by re-preparing her system in state $\pi^\alpha_j$ again, and then again forgetting $A_c$. 
Thus with the trivial choice $a_2=a'$, a subsequent measurement of $a_2$ will yield the same information as if the measurement 
of $a'$ had been performed instead of $a_1$. In other words, by the Definition \ref{defDis2}, this 
measurement of $a_1$ is non-disturbing.  

Now to prove the reverse implication, assume instead that all observables $\hat a$ can be measured 
without Bohr-disturbing the system $\beta$. For a quantum system, this means that any $\hat a$ can be measured 
in a way that (once the result is forgotten) the joint system is left in the same state $\varrho_c$. 
This is only possible if the algebra of observables on $\alpha$ is effectively classical, 
with no non-commuting structure. The state (\ref{nodisc}) is simply the way to write this classicality 
in quantum terms. Thus it must be the case that the joint state is Alice-concordant, with $\delta(\beta|c)=0$. 
Rewriting (\ref{nfr}) in positive terms, we can thus state 
\begin{theorem}
$$\forall \ \phi \in Q, \ \left[ \Disc(\beta| c) \iff \exists\ \hat{a}  :  \forall\ a \in {\mathbb S}_{\hat a|c}, \ \Dis^{\rm Bohr}(\beta|a,c)  
 \right]  $$
\end{theorem}
That is, if and only if the joint state is Alice-discordant (with $\delta(\beta|c)>0$) then 
there is some $\alpha$-observable such that the measurement thereof necessarily disturbs Bob's system $\beta$. 

The discord $\delta(\beta|a,c)$ as defined in \erf{defdisa} can be thought of as the minimum amount of quantum 
information about $\beta$ that is irrevocably lost when Alice  performs a measurement $a$. The loss occurs precisely 
because by choosing a particular measurement, Alice is eliminating 
``possible types of predictions regarding the future behaviour of the system''. 
The discord $\delta(\beta|c)$ is in fact a {\em quantification}, in bits, of Bohr's non-mechanical disturbance, 
the extent to which the measurement $a$ influences ``the very conditions which define the possible types of predictions''. 

It is the {\em complementarity} of different observables, the fact that they ``demand mutually exclusive 
experimental arrangements'' \cite{Boh35}, that  is the essence of the phenomenon of discord, and makes  
Bohr's notion of non-mechanical disturbance non-trivial. 
The trivial aspect to Bohr's notion is that any measurement $a$ can be disturbing, according to the Definition  \ref{defDis2}: 
\blk simply by destroying the system $\alpha$ 
Alice will lose information about the distant system $\beta$. The  non-trivial  point is that, for a suitable state, 
complementarity can make this disturbance inevitable for at least one observable $\hat a$. In fact, 
it is a plausible conjecture that for a discordant state disturbance is inevitable for {\em all} observables $\hat a$ with non-degenerate eigenvalues. 

The role of complementarity in classifying bipartite quantum states has previously been discussed in 
Refs.~\cite{Luo08,WuPouMol09}. However both of these papers considered the question of whether 
complementarity could play any role in the system at all, and concluded that it did not only when the system
was of the form 
\beq \label{consonant}
\varrho_c = \sum_j \wp_c(j) \ \pi^\alpha_j \otimes \pi^\beta_j,
\eeq where  
the $\pi^\beta_j$ are also mutually orthogonal rank-one projectors. These are simply states that 
are both Alice-concordant and Bob-concordant and may be called consonant states 
(in that they have zero {\em dissonance} \cite{Modi10}). While these conclusions are certainly valid, 
in the context of understanding Bohr's notion of disturbance it is more enlightening to concentrate
upon the asymmetric notion of discord, just as for understanding the EPR phenomenon it is useful 
to introduce the asymmetric notion of steering. The reason is the same: both EPR and Bohr were 
interested in the consequences, for one system (Bob's), of making a measurement upon a distant second system (Alice's).

Although not central to the discussion, 
it is worth highlighting the hierarchy of the nonlocality concepts relating to bipartite states 
which have been mentioned in this paper:
\beq
\textrm{Bell-nonlocality} \implies \textrm{steering} \implies \textrm{entanglement} \implies \textrm{discord} \implies \textrm{dissonance} \nonumber
\eeq
Note that for the asymmetric concepts (steering and discord), the implications apply regardless of which party we 
specify it for. Note also that all of these implications are strict (the converse implications are false).  Finally, 
returning to the point at the end of Sec.~\ref{sec:Ent}, it is for the case of (non-product) 
consonant states that there exists a (non-trivial) element of physical reality 
$\hat b$ by the EPR Criterion \ref{EPRcriterion}, whether by EPR's or Bohr's notion of disturbance. This is because such states 
have perfect correlations in the bases used in \erf{consonant} and yet, being concordant, there is no disturbance \`a la Bohr.  

\section{Discussion} \label{sec:Discussion}

I have suggested that both the EPR argument, and Bohr's counter-argument, can be better appreciated by attempting 
a rigorous formulation. In particular, my formalization allows us to see that both were correct, and that  their 
differing conclusions about the completeness of quantum mechanics hinged upon 
a differing conception of disturbance. For EPR, as long as two quantum systems no longer interact,   
a measurement performed by Alice's has no measurable effect on Bob's potential outcomes, and so does not  
disturb Bob's system. For Bohr, if Alice's measurement alters
the possible predictions she could otherwise have made regarding Bob's potential outcomes then that amounts 
to a disturbance of Bob's sytem. I have shown that Bohr's concept is intimately related to the concept 
of quantum discord, in that if and only if the state they share has no Alice-discord, there is a way for 
her to measure any observable in a way that entails no disturbance of Bob's system, in Bohr's sense.

As originally introduced, discord was a quantitative measure of the necessary loss of quantum 
information about Bob's system when Alice performs a measurement. Such a quantification is only 
possible when an information theory exists. This is the case for classical probabilities and for quantum states 
and measurements. For more general theories, such as Spekkens' toy-bit theory \cite{Spe07}, and 
the PR-box theory \cite{PopRoh94}, this is not necessarily the case. In this context, the present work 
shows that it is still possible to consider the presence or absence of discord by using the operational 
notion of Bohr-disturbance. In particular, in toy-bit theory,  at least for the case of 
a few toy-bits each held by Alice and Bob,  it is not possible to 
exhibit disturbance in Bohr's sense for LOCC-preparable bipartite states (that is, states that can be prepared 
from uncorrelated states by local operations and classical 
communication --- here via a toy-bit channel which decoheres in a particular basis, so to speak --- between Alice and Bob). 

 Returning now to quantum mechanics, the situation is unlike that of the toy theory, as there 
definitely exist non-entangled (and thus LOCC-preparable) states 
that are discordant. As I have argued in Sec.~\ref{sec:Ent}, entanglement is a necessary property, and 
pure entanglement a sufficient property, of quantum states for a proof of incompleteness using EPR's definitions. 
By contrast, discord is the necessary and sufficient phenomenon that underpins Bohr's counter-argument. 
Thus 
there is no evidence in Bohr's reply 
that he engaged with the EPR phenomenon of steering, nor with the concept of entanglement which it 
relies upon.

EPR's result (Theorem~\ref{thm:EPR}) states that orthodox quantum theory is incomplete, by EPR's necessary 
criterion for completeness that all elements of physical reality are represented in the theory. 
Now the existence of elements of physical reality is an operational concept for EPR, 
and for maximally entangled states \cite{Nielsen00a} all properties of $\beta$ would be elements of physical 
reality by this definition, because Alice can predict any property of $\beta$ by measuring her system 
 in the appropriate basis. Likewise for such states all properties of $\alpha$ would be elements of physical 
reality. Thus, in a complete theory, these should
be represented, by having values determined by the theory, independent of measurements 
made on the other system  --- see Sec.~\ref{sec:EPRdefs}. 
The impossibility of such a \grn local \blk representation, in {\em any} theory, for the correlations of maximally entangled qubit states, 
was exactly what Bell proved in 1964 \cite{Bel64}. 
That is, Bell's 1964 theorem can be stated, in precisely EPR's terms, as:
\begin{theorem}[Bell, 1964] $$\exists\   \phi \in Q : \forall\ \theta \in \Theta_\phi, \ 
\neg \Com^{\rm EPR}(\theta).$$ \label{thm:Bell}
\end{theorem}
Here I use the EPR superscript on $\Com^{\rm EPR}(\theta)$ to emphasize that this is 
EPR's notion of completeness, and in particular uses EPR's criterion for disturbance, not Bohr's. 

If Bell's theorem had been intuited by Bohr then it would of course have been the perfect 
response to the EPR paper: ``Your criterion for completeness implies that no theory for 
quantum phenomena can provide a complete description. 
So you must admit that either your concept of completeness is misguided, 
or that your conviction that `such a theory is possible' \cite{EPR35} is mistaken.'' 
The present work militates against the plausibility of an alternate history such as this 
(or that in Ref.~\cite{CavWis12}). The structure of Bohr's response to EPR is in no way anticipatory of Bell-nonlocality. 
Rather, Bohr's argument relied upon the phenomenon of quantum discord, a notion 
of quantum correlations with no requirement for entanglement, or steering, let alone Bell-nonlocality. 
It is no coincidence that \sch, who coined the terms entanglement and steering, and 
Bell, whose theorem overshadows all of the 1935 papers, were followers of Einstein, 
not Bohr. 

\section*{Acknowledgements}

I am grateful to Eric Cavalcanti for long and far-ranging discussions,  
to Terry Rudolph for stimulating remarks, 
and to Eric Cavalcanti, Rob Spekkens  and Travis Norsen  for detailed criticisms of the manuscript. 


\bibliography{../../QMCrefsPLUS.bib}

 \end{document}